\documentclass[a4paper]{revtex4}
\usepackage{graphicx}
\usepackage{fancyhdr}
\usepackage{amsmath}
\begin{document}

%Title of paper
\title{From Higgs measurements to constraints on new physics with Lilith}

% Repeat the \author .. \affiliation  etc. as needed
%
% \affiliation command applies to all authors since the last
% \affiliation command. The \affiliation command should follow the
% other information

\author{J. Bernon}
\affiliation{Laboratoire de Physique Subatomique et de Cosmologie, Universit\'e Grenoble-Alpes, CNRS/IN2P3, 53 Avenue des Martyrs, F-38026 Grenoble, France}
\author{B. Dumont}
\affiliation{Center for Theoretical Physics of the Universe, Institute for Basic Science (IBS), Daejeon 305-811, Republic of Korea}

\begin{abstract}
The properties of the observed Higgs boson with mass around 125~GeV are constrained by a
 wealth of experimental results targeting different combinations for the production and decay of a Higgs boson. 
In order to assess the compatibility of a non-Standard Model-like Higgs boson with all available results, we present {\tt Lilith}, a new public tool that makes use of 
signal strength measurements performed at the LHC and the Tevatron.
\end{abstract}

%\maketitle must follow title, authors, abstract
\maketitle

\thispagestyle{fancy}

% body of paper here - Use proper section commands
% References should be done using the \cite, \ref, and \label commands
% Put \label in argument of \section for cross-referencing
%\section{\label{}}

%%%%%%%%%%%%%%%%%%%%%%%%%%%%%%%%%%
\section{Introduction}
The discovery of a new particle with properties compatible with those of the Standard Model (SM) Higgs boson and mass around 125~GeV at CERN's Large Hadron Collider (LHC)~\cite{Aad:2012tfa, Chatrchyan:2012ufa} was a major breakthrough.
Detailed information on the properties of this Higgs boson was obtained from the measurements performed during Run~I of the LHC at 7--8~TeV center-of-mass energy~\cite{ATLAS-CONF-2015-007,Khachatryan:2014jba}, making it possible to test a large variety of models of new physics.
The results of the Higgs searches at the LHC are usually given in terms of signal strengths, $\mu$, which scale the number of signal events expected for the SM Higgs boson. The signal strength measurements can be combined and used to constrain new physics models in which {\it i)} the signal is a sum of the processes that exist for the SM Higgs boson, and {\it ii)} the Higgs couplings are defined through a simple rescaling of the corresponding SM ones.
In this talk we present a new public tool for performing a global fit to all available signal strength measurements, {\tt Lilith}~\cite{Bernon:2015hsa}. It is a library written in {\tt Python} that can easily be used in any {\tt Python} script as well as in {\tt C} and {\tt C++}/{\tt ROOT} codes, and for which we also provide a command-line interface.
As a primary input, it uses signal strengths results for which the fundamental production and decay modes have been unfolded from experimental categories. 
The experimental results are stored in {\tt XML} files, making it easy to modify and extend. The user input can be given in terms of reduced couplings or signal strengths, and is also specified in an {\tt XML} format.

%%%%%%%%%%%%%%%%%%%%%%%%%%%%%%%%%%
\section{Likelihoods and parametrization of new physics}
\begin{figure}[t]
\centering \includegraphics[scale=0.35]{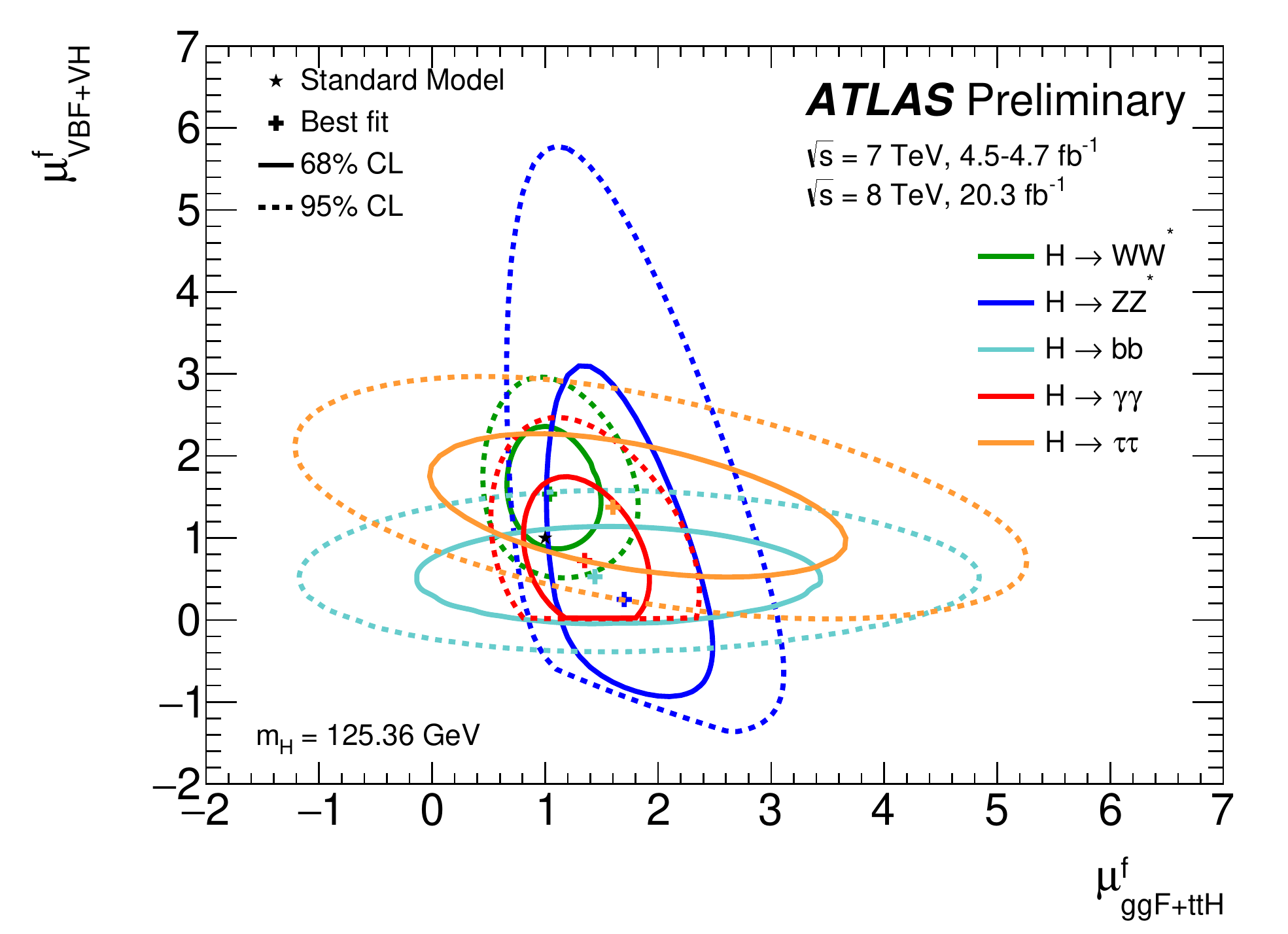} \includegraphics[scale=0.27]{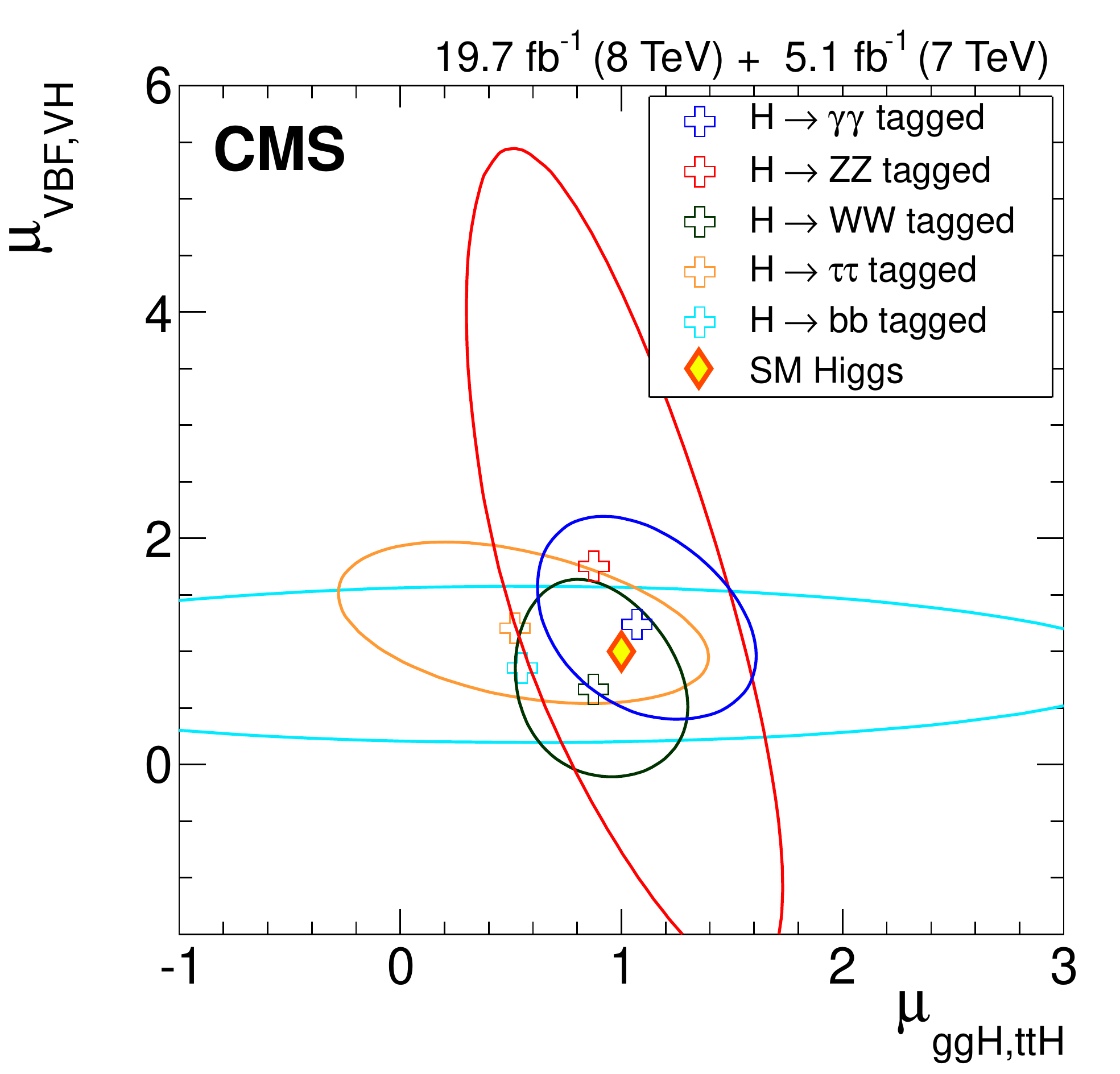}
\caption{Two dimensional ATLAS (left) and CMS (right) results in which the fundamental production modes are unfolded from experimental categories, in the plane $(\mu({\rm ggH+ttH}, Y), \mu({\rm VBF+VH}, Y))$ for $Y\in(\gamma\gamma$, $ZZ^*$, $WW^*$, $\tau\tau)$~\cite{ATLAS-CONF-2015-007,Khachatryan:2014jba}.}
\label{fig:results2Dmu}
\end{figure}

Each event category (corresponding to a given set of selection criteria) is sensitive to different combinations for production and decay of the SM Higgs boson, with various efficiency factors. Given this information, signal strength measurements can be combined and re-expressed after unfolding of the fundamental production and decay modes as
\begin{equation}
	\mu(X,Y) \equiv \frac{\sigma(X){\cal B}(H\to Y)}{\sigma^{\rm SM}(X){\cal B}^{\rm SM}(H\to Y)} \,,
	\label{eq:signalstr}
\end{equation}
for the different production modes $X\in({\rm ggH}, {\rm VBF}, {\rm VH=WH+ZH}, {\rm ttH})$ and decay modes $Y\in(\gamma\gamma$, $ZZ^*$, $WW^*$, $b\bar{b}$, $\tau\tau$, $\ldots)$ of the SM Higgs boson.
Results have been systematically given in terms of $\mu(X,Y)$ for the main decay modes probed by the ATLAS and CMS collaborations. In particular, results were often published in the plane $(\mu({\rm ggH+ttH}, Y), \mu({\rm VBF+VH}, Y))$, see Fig.~\ref{fig:results2Dmu}.

This information is taken into account in {\tt Lilith} in the following way: for two (combination of) production and decay processes $(X,Y)$ and $(X',Y')$, in the Gaussian approximation the experimental likelihood reads
\begin{equation}
- 2 \log L(\boldsymbol{\mu}) =
(\boldsymbol{\mu} - \hat{\boldsymbol{\mu}})^T
C^{-1}
(\boldsymbol{\mu} - \hat{\boldsymbol{\mu}}) \,, \label{eq:mu2d}
\end{equation}
where $\boldsymbol{\mu}^T = \left( \mu(X, Y) , \mu(X', Y') \right)$ and $C^{-1} = \begin{pmatrix} a&b\\ b&c \end{pmatrix}$ is the inverse of the covariance matrix. 
The list of experimental results present in the database of {\tt Lilith} can be found in~\cite{Bernon:2015hsa} (see also~\cite{LilithWebsite} for updates). It includes the results from direct searches for invisible decays of the Higgs boson through VBF and ZH production.
Whenever available, full 1- or 2-dimensional likelihood functions are used as they make it possible to go beyond the Gaussian approximation. All results are combined into a global likelihood function $L$ defined as the product of the individual likelihood functions.

Results given in terms of signal strengths, as in Eq.~\eqref{eq:signalstr}, can be matched to new physics scenarios with the introduction of factors $C_X$ and $C_Y$ that scale the amplitudes for the production and decay of the SM Higgs boson, respectively, as follows:
\begin{equation}
\mu(X,Y) = \frac{C_X^2 C_Y^2}{\sum_Y C_Y^2 {\cal B}^{\rm SM}(H\to Y)} 
\,. \label{eq:signalstr2}
\end{equation}
The factors $C_X$ and $C_Y$ can be identified to (or derived from) reduced couplings appearing in an effective Lagrangian~\cite{LHCHiggsCrossSectionWorkingGroup:2012nn,Heinemeyer:2013tqa}.
In particular, the factors $C_{\rm ggH}$ and $C_{\gamma\gamma}$---scaling gluon fusion and the decay into two photons, respectively---can be computed from the reduced couplings at LO or NLO QCD accuracy in {\tt Lilith}. It is also possible to specify decays into invisible and/or undetected particles. Given this information, {\tt Lilith} computes the signal strengths (alternatively they can   be given directly as input) and evaluates the global likelihood function. Contraints can then be put on new physics scenarios in a frequentist or Bayesian approach.

%%%%%%%%%%%%%%%%%%%%%%%%%%%%%%%%%%
\section{Quick introduction to Lilith}

{\tt Lilith} is a library written in {\tt Python} for constraining models of new physics against the
LHC results. The code is distributed under the GNU General Public License v3.0.
All necessary information on how to download and install it can be found on its official website~\cite{LilithWebsite}.
{\tt Lilith} requires {\tt Python~2.6} or more recent, but not the {\tt 3.X}~series.
The standard {\tt Python} scientific libraries, {\tt SciPy} and {\tt NumPy}, should furthermore be installed.

{\tt Lilith} can be used in three different ways:
\begin{enumerate} \setlength{\itemsep}{0pt} %\setlength{\parskip}{0pt}
\item from a a {\tt Python} code or an interactive session of {\tt Python}, using the {\tt Lilith} application programming interface (API);
\item calling the command-line interface (CLI) {\tt run\_lilith.py} from a shell;
\item or using the {\tt Lilith} interface to {\tt C} and {\tt C++}/{\tt ROOT}. 
\end{enumerate}

All three methods are documented in~\cite{Bernon:2015hsa}. A minimal example of use of the API is now presented:
\begin{verbatim}
 from lilith import *
 lcal = Lilith()
 lcal.readexpinput()
 lcal.readuserinputfile('userinput/example_mu.xml')
 lcal.computelikelihood()
 print '-2log(likelihood) =', lcal.l
\end{verbatim}
After importing the {\tt Lilith} library, an object {\tt lcal} is instantiated. The list of experimental results as well as the user input in {\tt XML} format are then read with the methods {\tt readexpinput} and {\tt readuserinputfile}, respectively. Finally, the likelihood is evaluated with the method {\tt computelikelihood} and the result is printed on the screen.

The {\tt XML} user input file consists in either reduced couplings or signal strengths (possibly with additional invisible or undetected decays) for one or several states contributing to the signal. The root tag should be {\tt <lilithinput>}, and is followed by either {\tt <reducedcouplings>} or {\tt <signalstrengths>}. For instance, scaling factors associated with the $H\to b\bar{b}$ and $H\to\gamma\gamma$ processes can be specified as follows,
\begin{verbatim}
<C to="bb">1.2</C>
<C to="gammagamma">0.8</C>
\end{verbatim}
The complete list of available attributes can be found in the manual~\cite{Bernon:2015hsa}.\footnote{In the latest release to date, {\tt Lilith-1.1.2}, a reduced coupling for $H\to\mu\mu$ is introduced and taken into account, but is not documented in the manual.} Alternatively, the signal strengths can be provided as input. For instance, taking $\mu({\rm ggH}, WW^*)$ and $\mu({\rm ttH}, b\bar{b})$ it would read
\begin{verbatim}
<mu prod="ggH" decay="WW">1.3</mu>
<mu prod="ttH" decay="bb">1.1</mu>
\end{verbatim}
Finally, with reduced couplings a {\tt <precision>} tag can be given, with possible values being {\tt "LO"} or {\tt "BEST-QCD"} for the computation of loop-induced couplings at LO or NLO QCD accuracy, respectively.

We also provide a CLI as well as an interface to {\tt C} and {\tt C++}/{\tt ROOT} in order to easily use {\tt Lilith} outside of a {\tt Python} script. All details and documentation are given in the manual~\cite{Bernon:2015hsa}.

%%%%%%%%%%%%%%%%%%%%%%%%%%%%%%%%%%
\section{Validation and example}

\begin{figure}[t]
\centering \includegraphics[scale=0.26]{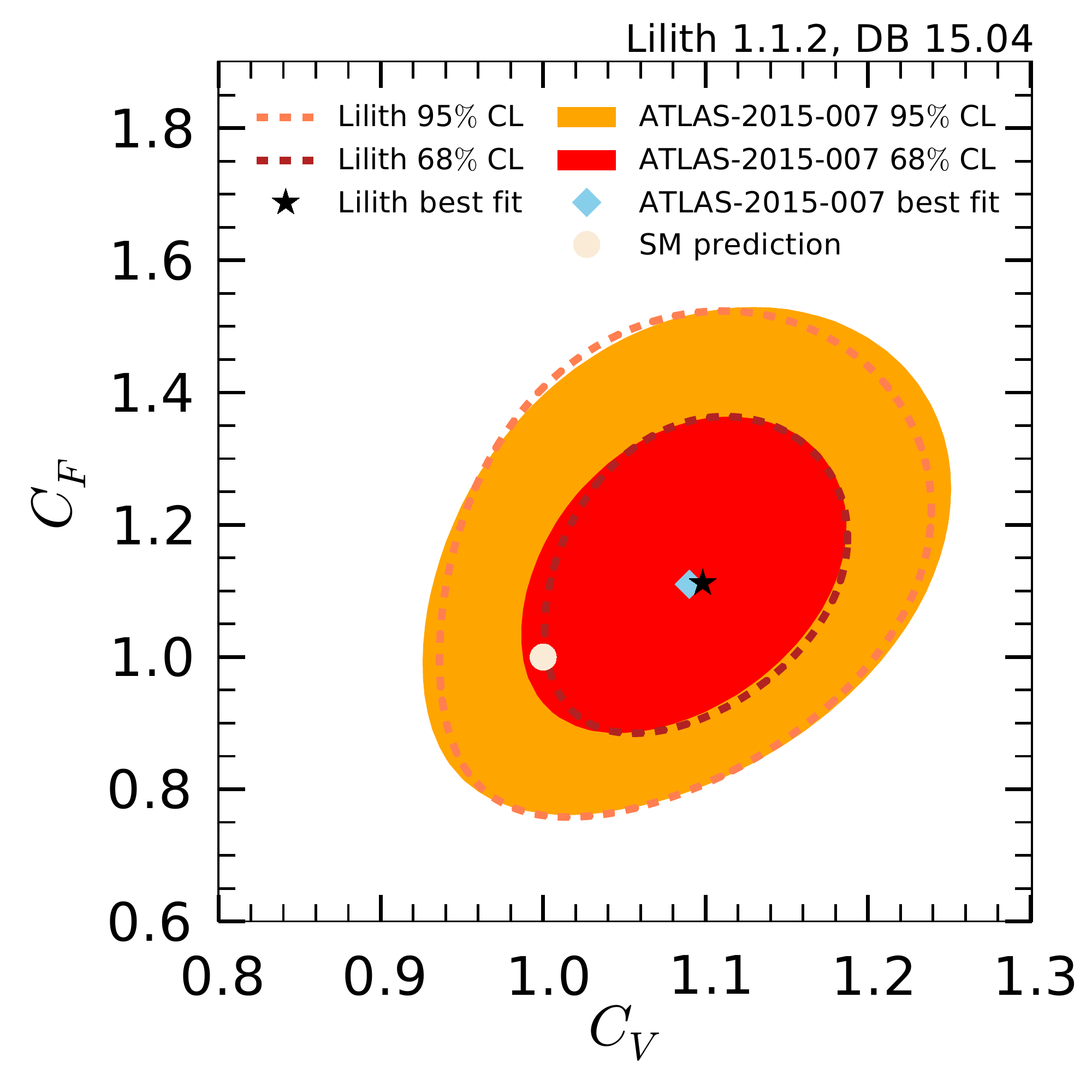} \includegraphics[scale=0.26]{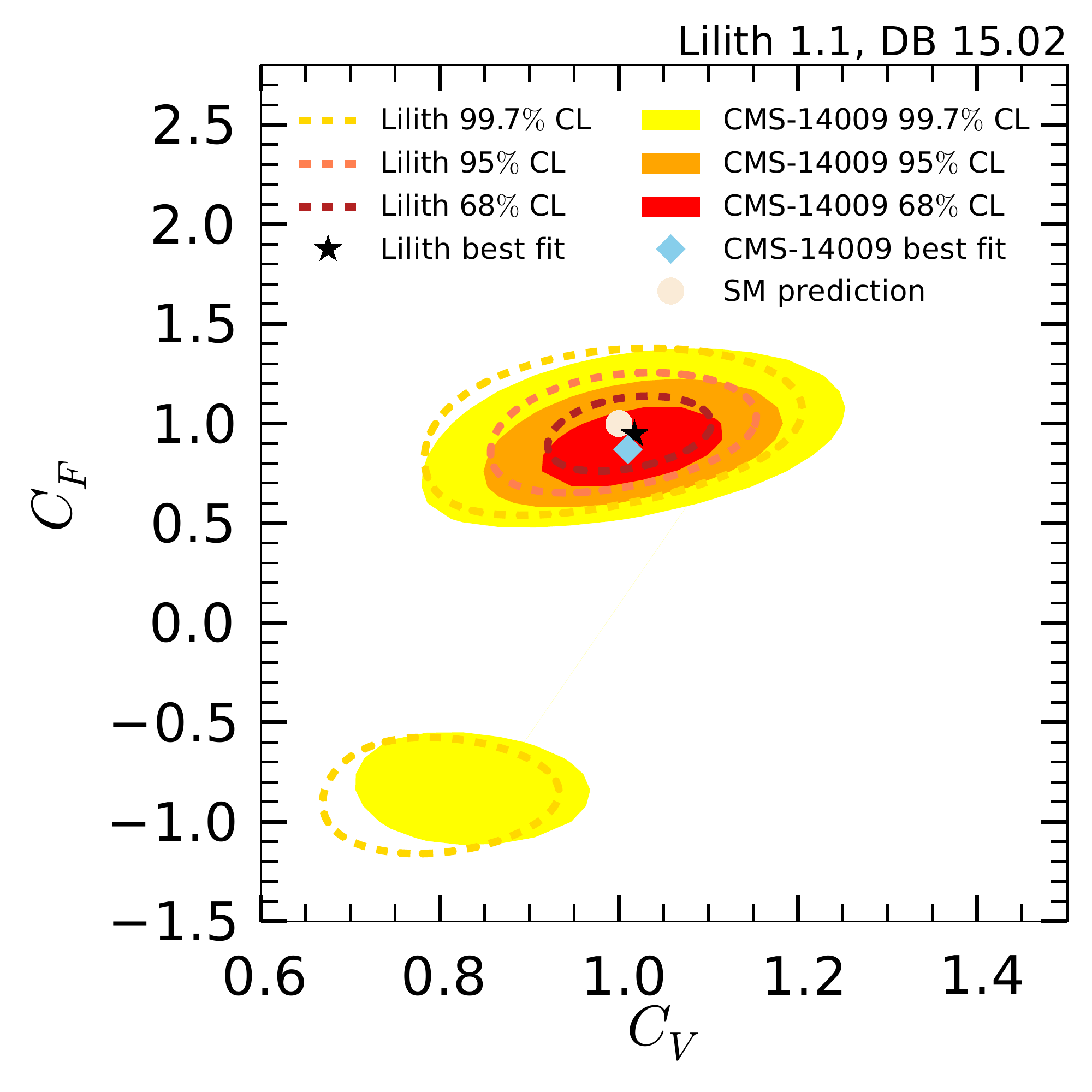}
\caption{Fits to the benchmark scenario $(C_V,C_F)$ using data from the ATLAS (left) and CMS (right) collaborations~\cite{ATLAS-CONF-2015-007,Khachatryan:2014jba}.
The red and orange filled surfaces correspond to the 68\% and 95\%~CL regions obtained by the ATLAS and CMS collaborations while the corresponding dashed lines show the {\tt Lilith} results. The black star indicates the position of the {\tt Lilith} best-fit point, the blue diamond is the ATLAS or CMS best-fit point and the white circle shows the SM prediction.}
\label{fig:validationCVCF}
\end{figure}

The Higgs likelihood used in {\tt Lilith} should be validated against official LHC results as approximations are made when reconstructing the full likelihood from 1- or 2-dimensional results. To this aim, we reproduce global fits to reduced couplings in various scenarios considered by ATLAS and CMS. An example is given in Fig.~\ref{fig:validationCVCF}. It shows very good agreement with the results from both experimental collaborations. More examples can be found in the manual~\cite{Bernon:2015hsa}.

\begin{figure}
\centering \includegraphics[scale=0.25]{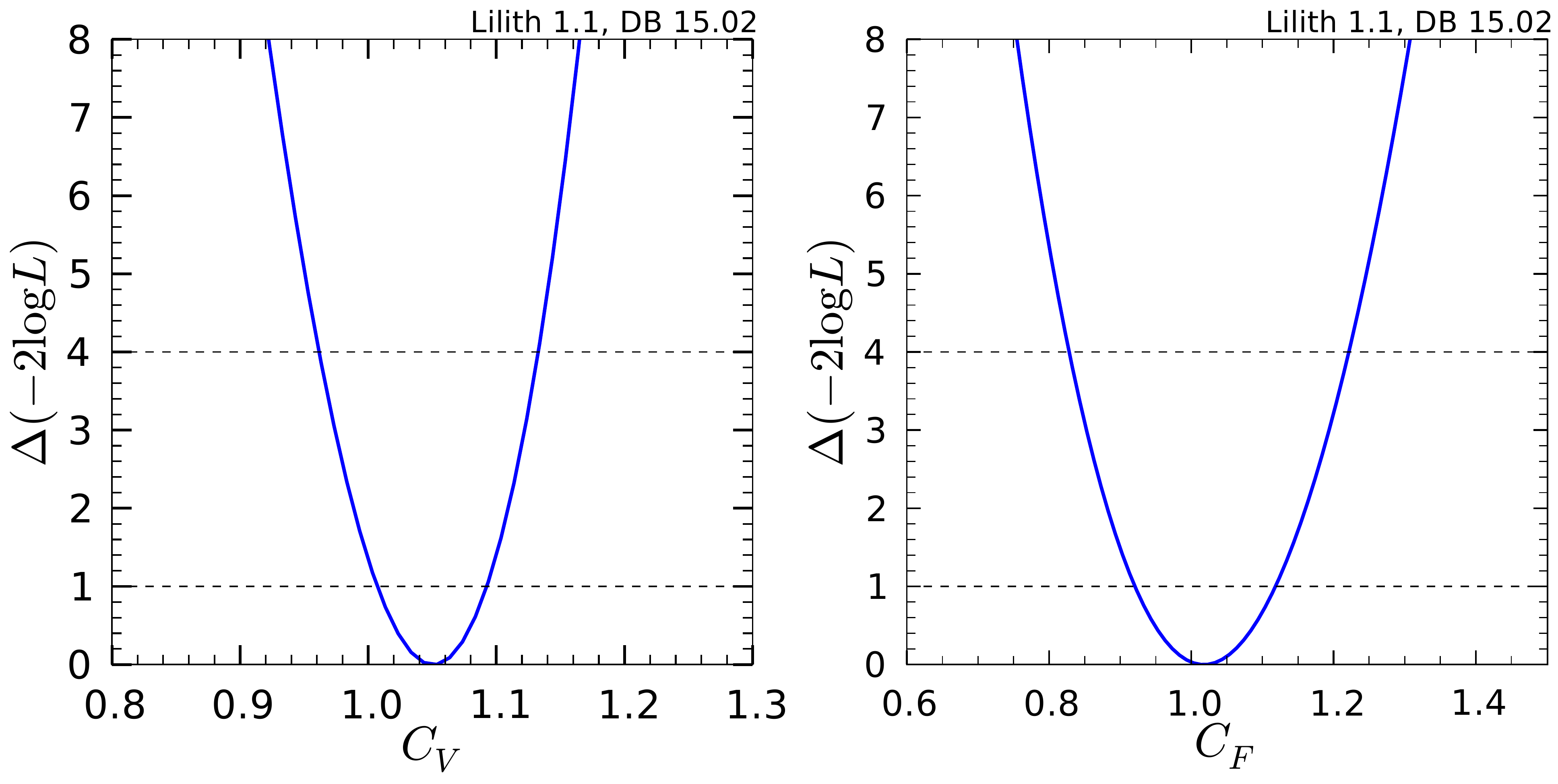} 
\caption{1-dimensional likelihood profiles of $C_V$~(left) and $C_F$~(right) from a global fit of the benchmark scenario $(C_V,C_F)$.}
\label{fig:CVCFglobalfit}
\end{figure}

Several examples are shipped with {\tt Lilith} (for the complete list, see~\cite{Bernon:2015hsa}). In the {\tt Python} code {\tt CVCF\_1dprofile.py}, 1-dimensional constraints on $C_V$ and $C_F$ are derived from a global fit to ATLAS and CMS results in the $(C_V,C_F)$ benchmark scenario. Results are shown in Fig.~\ref{fig:CVCFglobalfit}.
This example uses the library \texttt{iminuit}~\cite{iminuit}, a \texttt{Python} implementation of the \texttt{MINUIT}~\cite{James:1975dr} library, in order to minimize $-2\log L$ and derive the profiled likelihood around the minimum. The plotting library \texttt{matplotlib}~\cite{matplotlib} is used to produce the figures.

%%%%%%%%%%%%%%%%%%%%%%%%%%%%%%%%%%
\section{Conclusions}
Using all available information from the ATLAS and CMS collaborations to construct a likelihood is a non-trivial task. 
To this aim, we provide a new public tool, {\tt Lilith}.
{\tt Lilith} is a library written in {\tt Python}, and for which we provide an API as well as a command-line interface and a basic interface to {\tt C} and {\tt C++}/{\tt ROOT}.
The experimental results are read from a database in {\tt XML} format that is shipped with the code and which is easy to modify and extend. {\tt Lilith} uses as a primary input results in which the fundamental production and decay modes are unfolded from experimental categories.

New physics can be parametrized in terms of reduced couplings, or signal strengths directly, which are given as input to {\tt Lilith} in {\tt XML} format. If needed, scaling factors for the loop-induced processes and VBF production are computed taking into account QCD corrections. 
The likelihood is evaluated from a set of experimental results and given as output.
The Higgs likelihood of {\tt Lilith} obtained from the latest measurements at the LHC has been validated against ATLAS and CMS results and can be used to constrain new physics.

% If you have acknowledgments, this puts in the proper section head.
%\bigskip % extra skip inserted
%%%%%%%%%%%%%%%%%%%%%%%%%%%%%%%%%%
\begin{acknowledgments}
We thank the organizers of HPNP2015 for setting up this nice conference, and Sabine Kraml for her support and useful discussions.
This work was supported in part by the ANR project DMAstroLHC, the
``Investissements d'avenir, Labex ENIGMASS'', and by the IBS under
Project Code IBS-R018-D1.
\end{acknowledgments}

\bigskip % extra skip inserted
% Create the reference section using BibTeX:
%\bibliography{basename of .bib file}

\end{document}